# Predicting Lymph Node Metastasis in Head and Neck Cancer by Combining Many-objective Radiomics and 3-dimensioal Convolutional Neural Network through Evidential Reasoning*

Zhiguo Zhou, Liyuan Chen, David Sher, Qiongwen Zhang, Jennifer Shah, Nhat-Long Pham, Steve Jiang, and Jing Wang

*Abstract*— Lymph node metastasis (LNM) is a significant prognostic factor in patients with head and neck cancer, and the ability to predict it accurately is essential for treatment optimization. PET and CT imaging are routinely used for LNM identification. However, uncertainties of LNM always exist especially for small size or reactive nodes. Radiomics and deep learning are the two preferred imaging-based strategies for node malignancy prediction. Radiomics models are built based on handcrafted features, and deep learning can learn the features automatically. We proposed a hybrid predictive model that combines many-objective radiomics (MO-radiomics) and 3-dimensional convolutional neural network (3D-CNN) through evidential reasoning (ER) approach. To build a more reliable model, we proposed a new many-objective radiomics model. Meanwhile, we designed a 3D-CNN that fully utilizes spatial contextual information. Finally, the outputs were fused through the ER approach. To study the predictability of the two modalities, three models were built for PET, CT, and PET&CT. The results showed that the model performed best when the two modalities were combined. Moreover, we showed that the quantitative results obtained from the hybrid model were better than those obtained from MO-radiomics and 3D-CNN.

## I. INTRODUCTION

Head and neck (H&N) cancer is the sixth most common cancer type worldwide [1]. Since one third of the body's lymph nodes are located in the head and neck, lymph node metastases (LNM) is one of the most important metastatic ways for malignant tumors and are a significant prognostic factor for patients with H&N cancer [2]. Clinically, CT and PET have been routinely used for LNM identification. CT is used commonly for detection purposes, and PET has the great advantages of location and histological nature of the LNM [3]. Although some lymph nodes are clearly positive due to their large size or high activity on PET-CT, there is often significant uncertainty about the malignant potential of lymph nodes in H&N cancer.

Imaging-based classification is solved by two major strategies: handcrafted feature-based models and feature learning-based models. Among the handcrafted feature models, radiomics appears to be a highly promising solution [4]. Defined as the extraction and analysis of a large number of quantitative features, radiomics has been applied successfully to solve various prediction problems, such as treatment outcome prediction [5], and survival analysis [6]. Huang et al. [7] developed a radiomics model to predict LNM in colorectal cancer. The features were extracted from CT images, and multivariable logistic regression was used to build the predictive model. To build a more reliable model, our group developed a multi-objective radiomics model [8] which considered both sensitivity and specificity as the objective functions. For feature learning-based models, deep learning is a powerful method that has been used to predict outcomes in cancer therapy. Sung et al. [9] described the use of deep learning methods such as the convolutional neural network (CNN), deep belief network, and stacked de-noising auto-encoder to predict lung nodule malignancy. Zhu et al. designed a new CNN model to predict survival in lung cancer [10]. Yang et al. [11] built a model that combined the recurrent neural network and multinomial hierarchical regression decoder to predict breast cancer metastasis.

As both handcrafted feature and feature learning models have yielded promising results, one challenge from a practical point of view is to determine which model is more suitable to predict LNM. Manually extracted features and automatic learned features are complementary [12], and therefore may yield more stable results. Therefore, a strategy that combines both handcrafted and learning models is a more suitable choice to predict LNM.

In this work, we proposed a hybrid model that combines the multi-objective radiomics and 3D-CNN through evidential reasoning (ER) to predict LNM in H&N cancer. Because the multi-objective model [8] can only handle binary problems, we proposed a new MO-radiomics model that can predict the three classes of lymph nodes- normal, suspicious, and involved. Other than using sensitivity and specificity as the objectives in multi-objective model, procedure accuracy (PA) and user accuracy (UA) in confusion matrix (CM) were considered simultaneously as objectives in the proposed model. We adopted CNN [13], one of the most effective feature learning models consisting of conventional, max pooling, and fully connected layers. Since the nodes appear in 3D in the images, a 3D-CNN model was designed to consider spatial contextual information. The final output was obtained by fusing MO-radiomics and 3D-CNN model outputs through the ER approach [14]. ER was originally proposed to deal with multiple attribute decision analysis problems with both qualitative and quantitative attributes under uncertainty. Also, ER can yield the more reliable fusing results.

*Research supported by the American Cancer Society (ACS-IRG-02-196) and the US National Institutes of Health (R01 EB020366).

Zhiguo Zhou, Liyuan Chen, David Sher, Jennifer Shah, Nhat-Long Pham, Steve Jiang, and Jing Wang are with the University of Texas Southwestern Medical Center, Dallas, TX 75390 USA (corresponding author to provide phone: 214-648-1795; fax: 214-648-7389; e-mail: Jing.Wang@ UTSouthwestern.edu). Zhiguo Zhou and Liyuan Chen contributed equally to this work.

Qiongwen Zhang is with the State Key Lab of Biotherapy and Cancer Center, West China Hospital, Sichuan University, Chengdu, China.

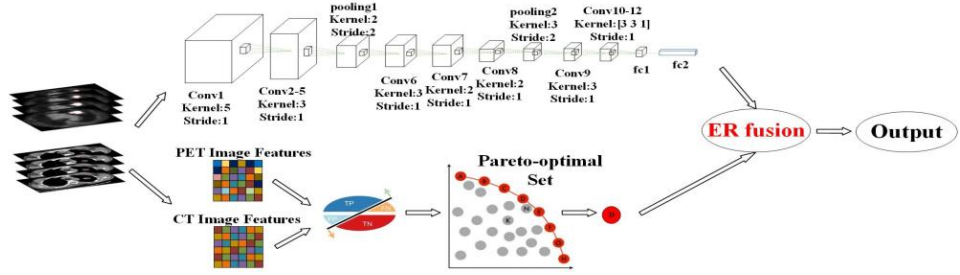

Figure 1. Workflow of the proposed hybrid model.

## II. MATERIALS AND METHODS

### A. Patient dataset

The study included 31 patients with H&N cancer who had enrolled in the involved field trial between 2016 and 2017 at UT Southwestern Medical Center. Pretreatment PET and CT images were exported from picture archiving and communication system. Nodal status was reviewed for all trial patients by a radiation oncologist and a nuclear medicine radiologist. Involved and suspicious nodes were contoured on contrast-enhanced CT under the guidance of PET. The lymph nodes for the first 21 patients were used for model training, including 53 involved nodes, 39 suspicious nodes, and 30 normal nodes. Then, the predictive model was validated on the remaining independent 10 patients with 13 involved nodes, 9 suspicious nodes, and 17 normal nodes. A total of 122 nodes were used for training and 39 were used for testing.

### B. Model overview

The workflow of the hybrid model is illustrated in Figure 1. First, the extracted nodes were input into the 3D-CNN and MO-radiomics model. Then, the two model outputs were fused by ER to obtain the final output.

### C. MO-radiomics model

In MO-radiomics, image features including intensity, texture, and geometric are extracted from the contoured lymph nodes (involved and suspicious) in PET and CT images. Additionally, at least one normal LN of similar size to the suspicious lymph nodes was contoured to train the predictive model for each patient. Intensity features include minimum, maximum, mean, standard deviation, sum, median, skewness, kurtosis, and variance. Geometry features include volume, major diameter, minor diameter, eccentricity, elongation orientation, bounding box volume, and perimeter. Texture features are based on 3D gray-level co-occurrence and are extracted as follows: energy, entropy, correlation, contrast, texture variance, sum-mean, inertia, cluster shade, cluster prominence, homogeneity, max-probability, and inverse variance. A total of 257 features were extracted for each PET and CT, respectively.

Then we used the support vector machine to build the predictive model with parameters denoted by $\alpha = \{\alpha_1, \cdots, \alpha_M\}$, where $M$ is the number of model parameters. All features including PET and CT imaging features are denoted by $\beta = \{\beta_1, \cdots, \beta_N\}$, where $N$ is the number of features. PAs and UAs in CM were taken as objective functions because of the three classes of lymph nodes. We maximized $f_{PA}^i$ and $f_{UA}^i$ simultaneously to obtain the Pareto-optimal set:

$$f = \max_{\alpha,\beta}(f_{PA}^i, f_{UA}^i, i = 1,2,3). \quad (1)$$

where $f_{PA}^i, f_{UA}^i$ represent the objective functions of PA and UA, respectively where $f_{PA}^i, f_{UA}^i$ represent the objective functions of PA and UA, respectively. According to equation (1), six objective functions were used. The final solution of the selected features and model parameters can be selected from the Pareto-optimal set according to different clinical needs.

To solve the optimization problem defined in equation (1), we developed a many-objective optimization algorithm based on our previous multi-objective algorithm [8]. The proposed algorithm also consists of two phases: (1) Pareto-optimal solution generation; (2) best solution selection. The first phase is same as that of the multi-objective algorithm [8], which includes initialization, clonal operation, mutation operation, deleting operation, population update, and termination detection. In the second phase, the final solution is selected according to accuracy and area under the curve (AUC). Assume that the thresholds for accuracy are denoted by $T_{acc}$. The Pareto-optimal solution is denoted by $D = \{D_1, D_2, \cdots, D_P\}$. The corresponding accuracy and AUC for each individual $D_i, i = 1, 2, \cdots, P$ are denoted by $D_i^{acc}, D_i^{AUC}, i = 1,2,\cdots,P$, respectively. The procedure to select the best solution is as follows: Step 1) For each solution set $D_i, i = 1,2,\cdots,P$, if $D_i^{acc} > T_{acc}$, $D_i$ is selected. All selected candidates constitute the new candidate set denoted by $D_C = \{D_C^1, D_C^2, \cdots, D_C^Q\}$, where $Q$ is the number of the selected individual, i.e., feasible solutions. Step 2) The final solution $P^*$ is selected with the highest AUC in $D_C$.

### D. 3D-CNN model

The architecture of the proposed 3D-CNN model includes 12 convolutional layers, 2 max pooling layers, and 2 fully connected layers (Figure 1). Each convolutional layer is equipped with rectified linear unit (ReLU) activation [15] and batch normalization. The construction order of different kinds of layers in the architecture is shown in Table I.

Since the max-pooling layer provides basic translation invariance to the internal representation, the convolutional and max-pooling layers are arranged alternately in the proposed architecture. In addition, with the help of the max-pooling layer which down-samples the feature maps, the convolutional layers in the architecture can capture both local and global features. In the proposed architecture, the first 8 convolutional layers are unpadded and the last 4 are padded to deeply extract and analyze features of the 3D image. To prevent back-propagated gradients from vanishing or exploding, we used Xavier initialization and categorical cross entropy loss function to train the proposed model. If we use CT or PET imaging only, the 3D-CNN input is a volumetric image. If we use PET and CT simultaneously, the input consists of two volumetric images, each serving as a channel of the final 4D data input. Data augmentation by rotating 3D images and Synthetic Minority Over-sampling technique [16] is used to balance and increase training samples. The network produces three probabilities for each class.

TABLE I. 3D-CNN ARCHITECTURE

| Layer | Kernel Size | Stride | Output Size | Features Volumes |
|---|---|---|---|---|
| Input | - | | 48×48×32 | 1(or 2) |
| C1 | 5×5×5 | [1 1 1] | 44×44×28 | 64 |
| C2 | 3×3×3 | [1 1 1] | 42×42×26 | 64 |
| C3 | 3×3×3 | [1 1 1] | 40×40×24 | 64 |
| C4 | 3×3×3 | [1 1 1] | 38×38×22 | 64 |
| C5 | 3×3×3 | [1 1 1] | 36×36×20 | 64 |
| MP1 | 2×2×2 | [2 2 2] | 18×18×10 | 64 |
| C6 | 3×3×3 | [1 1 1] | 16×16×8 | 64 |
| C7 | 2×2×2 | [1 1 1] | 15×15×5 | 64 |
| C8 | 3×3×3 | [1 1 1] | 13×13×4 | 64 |
| MP2 | 3×3×3 | [2 2 2] | 6×6×2 | 64 |
| C9 | 3×3×3 | [1 1 1] | 6×6×2 | 64 |
| C10 | 3×3×1 | [1 1 1] | 6×6×2 | 64 |
| C11 | 3×3×1 | [1 1 1] | 6×6×2 | 64 |
| C12 | 3×3×1 | [1 1 1] | 6×6×2 | 32 |
| FC1 | | | 1×1×1 | 256 |
| FC2 | | | 1×1×1 | 3 |

*C indicates Convolution layer + ReLU layer +Batch Normalization layer; MP indicates Max-pooling layer; GAP indicates Global-average-pooling layer and FC indicates Fully-connected layer.

*E. ER fusion*

After obtaining the outputs from the two models, the final output is generated using ER. Assume that $P^1 = \{p_1^1, p_2^1, p_3^1\}$ represents the output of MO-radiomics, and the 3D-CNN output is denoted by $P^2 = \{p_1^2, p_2^2, p_3^2\}$. They satisfy the following constraint:

$$\sum_{i=1}^{3} P_i^j = 1, \quad 0 \leq P_i^j \leq 1, j = 1,2, \quad (2)$$

Before using ER fusion, the weight $\omega = \{\omega_1, \omega_2\}$, which satisfies $\omega_1 + \omega_2 = 1, 0 \leq \omega_j \leq 1$, needs to be calculated. In this work, $\omega$ is calculated based on accuracy in the training stage. Assume that the accuracy for the two trained models is $A_j, j = 1,2, \omega_j$ is:

$$\omega_j = \frac{A_j}{\sum_{j=1}^{2} A_j}, j = 1,2, \quad (3)$$

The final output $P_i, i = 1,2,3$ is obtained through the following Equations:

$$P_i = \frac{\mu \times \left[\prod_{j=1}^{N}\left(\omega_j P_i^j + 1 - \omega_j \sum_{i=1}^{M} P_i^j\right) - \prod_{j=1}^{N}\left(1 - \omega_j \sum_{i=1}^{M} P_i^j\right)\right]}{1 - \mu \times \left[\prod_{j=1}^{N}(1-\omega_j)\right]}, \quad (4)$$

$$\mu = \left[\sum_{i=1}^{M} \prod_{j=1}^{N}\left(\omega_j P_i^j + 1 - \omega_j \sum_{i=1}^{M} P_i^j\right) - (N-1) \prod_{j=1}^{N}\left(1 - \omega_j \sum_{i=1}^{M} P_i^j\right)\right]^{-1}, \quad (5)$$

where $M = 3, N = 2$. Finally, the label $L$ is obtained by:

$$L = \max(P_i). \quad (6)$$

III. RESULTS

Besides combining PET and CT as input, we also used PET and CT alone to build the predictive models. Two typical radiomics (Radiomics) and CNN (XmaxNet) methods were also performed for comparison. For the many-objective training algorithm, the population number was set to 100, while the maximal generation number was set to 200. The mutation probability was set to 0.9 in the mutation operation. Five-cross-validation was used to train the model. Because three categories of nodules need to be handled, CM, accuracy, and multiclass AUC [17] were used to evaluate the performance. All the comparative results were statistical significance ($p$-value<0.01). The accuracy for five models under three situations is shown in Figure 2. Compared XmasNet and Radiomics with our proposed single and hybrid models, we demonstrated that the hybrid model can obtain the same or better accuracy. Although accuracy is the same as that of MO-radiomics and 3D-CNN in CT imaging, it improved in the hybrid model, indicating the effectiveness of ER fusion. The multiclass AUC shown in Figure 3 demonstrated that the hybrid model outperformed the four models, suggesting that results are more reliable after combination. On the other hand, improving the accuracy of LNM prediction in clinical use can predict overall survival and distant metastases more accurately.

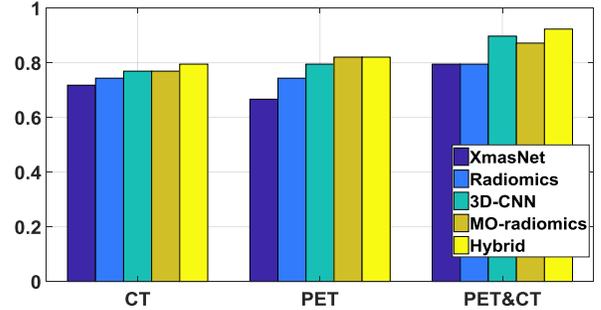

Figure 2. Accuracy for five models under three situations.

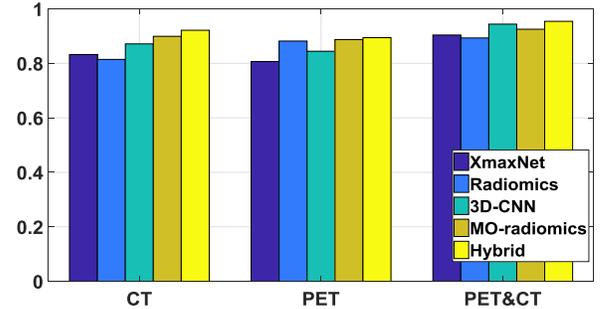

Figure 3. AUC for five models under three situations.

TABLE II. CONFUSION MATRIX FOR MO-RADIOMICS

| | Node | Predicted Normal | Predicted Suspicious | Predicted Involved | UA |
|---|---|---|---|---|---|
| CT | Normal | 17 | 0 | 0 | 1.00 |
| | suspicious | 3 | 2 | 4 | 0.22 |
| | involved | 1 | 1 | 11 | 0.85 |
| | PA | 0.89 | 0.67 | 0.73 | |
| PET | Normal | 15 | 2 | 0 | 0.88 |
| | suspicious | 1 | 6 | 2 | 0.67 |
| | involved | 2 | 0 | 11 | 0.85 |
| | PA | 0.83 | 0.75 | 0.85 | |
| PET & CT | Normal | 17 | 0 | 0 | 1.00 |
| | suspicious | 1 | 5 | 3 | 0.56 |
| | involved | 1 | 0 | 12 | 0.92 |
| | PA | 0.90 | 1.00 | 0.80 | |

TABLE III. CONFUSION MATRIX FOR 3D-CNN

| | Node | Predicted Normal | Predicted Suspicious | Predicted Involved | UA |
|---|---|---|---|---|---|
| CT | Normal | 14 | 3 | 0 | 0.82 |
| | suspicious | 2 | 5 | 2 | 0.56 |
| | involved | 2 | 0 | 11 | 0.86 |
| | PA | 0.78 | 0.63 | 0.85 | |
| PET | Normal | 17 | 0 | 0 | 1.00 |
| | suspicious | 0 | 3 | 6 | 0.33 |
| | involved | 2 | 0 | 11 | 0.85 |
| | PA | 0.90 | 1 | 0.65 | |
| PET & CT | Normal | 16 | 1 | 0 | 0.94 |
| | suspicious | 0 | 9 | 0 | 1.00 |
| | involved | 2 | 1 | 10 | 0.77 |
| | PA | 0.89 | 0.82 | 1.00 | |

CM results for the MO-radiomics model under three situations is shown in Table II, and those for 3D-CNN are shown in Table III. The CM for the proposed hybrid model is shown in table IV. All three CMs show that normal and involved nodes are easier to predict than suspicious nodes probably because suspicious and involved nodes are too similar to differentiate. When PET and CT are combined, better PA and UA can be obtained. Moreover, the hybrid model can yield the best PA and UA. Overall, our proposed predictive model is more effective than two single models in predicting LNM.

TABLE IV. CONFUSION MATRIX FOR HYBRID MODEL

| | Node | Predicted Normal | Predicted Suspicious | Predicted Involved | UA |
|---|---|---|---|---|---|
| CT | Normal | 16 | 1 | 0 | 0.94 |
| | suspicious | 2 | 4 | 3 | 0.44 |
| | involved | 1 | 1 | 11 | 0.85 |
| | PA | 0.84 | 0.67 | 0.79 | |
| PET | Normal | 16 | 1 | 0 | 0.94 |
| | suspicious | 1 | 5 | 3 | 0.56 |
| | involved | 2 | 0 | 11 | 0.85 |
| | PA | 0.84 | 0.83 | 0.79 | |
| PET & CT | Normal | 16 | 1 | 0 | 0.94 |
| | suspicious | 0 | 9 | 0 | 1 |
| | involved | 2 | 0 | 11 | 0.85 |
| | PA | 0.89 | 0.90 | 1.00 | |

IV. CONCLUSION

We proposed a hybrid model that predicts LNM in H&N cancer by combining MO-radiomics and 3D-CNN fused by the ER approach. To obtain more reliable performance, a new MO-radiomics model was developed based on our previous work, in which PAs and UAs in CM are considered as objective functions. Meanwhile, a 3D-CNN model was developed to make full use of contextual information in the images. The final output was obtained by combining the two model outputs using the ER approach. Our experimental results showed that the proposed model that combines PET and CT outperformed the two single models and classic methods.

In the MO-radiomics model, PAs and UAs are optimized simultaneously. In fact, these two types of objective functions can be trained alternately, which can potentially further improve the model performance. To obtain a more robust model, the transfer learning can be introduced into the 3D-CNN model as a next step. The dataset will also be expanded to include more patient data for building and validating the model so it can be applied successfully to clinical settings in the future.

ACKNOWLEDGMENT

The authors would like to thank Dr. Damiana Chiavolini for editing.